# Bottom-up approach to assess carbon emissions of battery electric vehicle operations in China

Hong Yuan[1], Minda Ma[2, 3*]

1 School of Management Science and Real Estate, Chongqing University, Chongqing, 400045, PR China

2 Building Technology and Urban Systems Division, Energy Technologies Area, Lawrence Berkeley National Laboratory, Berkeley, CA 94720, United States

3 School of Architecture and Urban Planning, Chongqing University, Chongqing, 400045, PR China

(Corresponding Author: maminda@lbl.gov; minda.ma@cqu.edu.cn)

**ABSTRACT**

The transportation sector is the third-largest global energy consumer and emitter, making it a focal point in the transition toward the net-zero future. To accelerate the decarbonization of passenger cars, this work is the first to propose a bottom-up charging demand model to estimate the operational electricity use and associated carbon emissions of best-selling battery electric vehicles (BEVs) in various climate zones in China during the 2020s. The findings reveal that (1) the operational energy demand of the top-20 selling BEV models in China, such as Tesla, Wuling Hongguang, and BYD, increased from 601 to 3054 giga-watt hours (GWh) during 2020-2022, with BEVs in South China contributing more than half of the total electricity demand; (2) from 2020 to 2022, the energy and carbon intensities of the best-selling models decreased from 1364 to 1095 kilowatt-hour per vehicle and from 797 to 621 kilograms of carbon dioxide ($CO_2$) per vehicle, respectively, with North China experiencing the highest intensity decline compared to that in other regions; and (3) the operational energy demand of BEV stocks in China increased from 4774 to 12,048 GWh during 2020-2022, while the carbon emissions of BEV stocks rose to 6.8 mega-tons of $CO_2$ in 2022, reflecting an annual growth rate of ~50%. In summary, this work delves into the examination and contrast of benchmark data on a nation-regional scale, as well as performance metrics related to BEV chargings. The primary aim is to support nationwide efforts in decarbonization, aiming for carbon mitigation and facilitating the swift evolution of passenger cars toward a carbon-neutral future.

**Keywords:** Battery electric vehicles, Charging demand, Bottom-up framework, Electricity use, Electrification and power decarbonization

**NONMENCLATURE**

| *Abbreviations* | |
|---|---|
| BEV | Battery electric vehicle |
| GWh | Gigawatt-hours |
| $MtCO_2$ | Mega-tons of carbon dioxide |
| NEDC | New European Driving Cycle |
| *Symbols* | |
| $E_k$ | Electricity consumption in the climate zone $k$ |
| $E$ | Total electricity consumption of top-20 selling BEV models in China |
| $Energy_i$ | Battery energy of version $i$ |
| $Mileage_k$ | Avenge mileage in the climate zone $k$ |
| $NEDC_i$ | NEDC of version $i$ |
| $Sales_j$ | Sales of model $j$ |
| $X$ | Total carbon emissions of top-20 selling BEV models in China |
| $\gamma_i$ | Sales share of version $i$ in the same model $j$ |
| $\eta$ | Emission factor of power sector |
| $\Theta_j$ | Electricity consumption of model $j$ |
| $\lambda_i$ | NEDC's degradation coefficient in the spring and autumn sessions of version i |
| $\upsilon_i$ | Electricity consumption in the summer and winter sessions of version $i$ |
| $\rho_i$ | NEDC's degradation coefficient in the summer and winter sessions of version i |
| $\omega_i$ | Electricity consumption in the spring and autumn sessions of version $i$ |
| $\Omega_i$ | Annual electricity consumption of version $i$ |

## 1. INTRODUCTION

The transportation sector is a noteworthy contributor to global fuel consumption and greenhouse gas emissions [1]. Accounting for approximately 50% of the total worldwide emissions of air pollutants, the transportation sector has emerged as a pivotal catalyst for urban air pollution [2]. Currently, electrification is regarded as one of the best practical solutions for decarbonizing road transportation, especially for passenger cars [3]. Battery electric vehicles (BEVs), in particular, exhibit high energy efficiency and lack contributions to local pollution, thereby presenting a pragmatic alternative to conventional internal combustion engine vehicles [4]. In particular, the widespread adoption of BEVs has the potential to significantly curtail petroleum consumption by 98% per kilometer and reduce fossil fuel usage by 25% to 50% [5]. Since 2015, China has consistently maintained its global dominance in the EV market, serving as a representative force in the realm of research and development pertaining to EVs [6]. At present, the aggregate volume of EVs in China is 294 million, accounting for 13.8% of the global EV market [7]. Consequently, a meticulous estimation of the charging demand and concomitant carbon emissions associated with BEVs in China is highly important for attaining carbon neutrality within the transportation sector and aligning with the objectives delineated in the Paris Climate Agreement. The charging demand of BEVs is intricately shaped by a myriad of factors, encompassing environmental temperature, traffic conditions, the stock of BEVs, the accessibility of charging infrastructure, and driver behavior [8]. The amalgamation of these factors to formulate a precise charging demand model represents a multifaceted undertaking. Currently, research endeavors aimed at estimating the charging demand for BEVs predominantly center on microlevel analyses, focusing on communities and cities [9, 10]. However, from a national or large-scale regional standpoint, there is a dearth of comprehensive studies that systematically evaluate the charging demand for BEVs, particularly within the realm of civilian charging [11-13].

To bridge this gap, this study is the first to propose a groundbreaking bottom-up framework [14-16] designed to evaluate charging demand within three distinct climate zones in China—North China, the Middle and Lower Reaches of the Yangtze River (MLRYR), and South China—focused specifically on the 2020s. The primary objective of this study is to assess the charging demand of each BEV model considering its various versions and considering parameters such as vehicle model performance, sales, mileage, energy intensity, and climate conditions. Furthermore, a bottom-up framework is employed to measure the actual electricity demand and associated carbon emissions of best-selling BEV model operations at both the national and regional levels. Additionally, the energy and emission intensities of best-selling BEV model operations, along with the energy demand and associated emissions of BEV stocks in China's passenger car sector, are analyzed.

## 2. MATERIAL AND METHODS

### 2.1 Bottom-up charging demand model of BEVs

Choosing the top-20 selling BEV models in terms of sales over a period of years, the annual electricity consumption of each BEV model and its different versions can be defined as follows:

$$\omega_i = \gamma_i \cdot Sales_j \cdot \frac{Mileage_k \cdot Energy_i}{\lambda_i \cdot NEDC_i} \quad (1)$$

$$v_i = \gamma_i \cdot Sales_j \cdot \frac{Mileage_k \cdot Energy_i}{\rho_i \cdot NEDC_i} \quad (2)$$

Therefore, the annual electricity consumption of version *i* can be expressed as follows:

$$\Omega_i = \omega_i + v_i, \quad (3)$$

Hence, the annual electricity consumption of model *j* can be defined as follows:

$$\Theta_j = \sum_{i=1}^{l} \Omega_i \quad (4)$$

Subsequently, the annual electricity consumption across different climate zones is defined as follows:

$$E_k = \sum_{j=1}^{m} \Theta_j \quad (5)$$

Finally, the total electricity consumption of best-selling BEV model operations in China (*E*) can be given as follows:

$$E = \sum_{k=1}^{n} E_k \quad (6)$$

### 2.2 Bottom-up emission model of BEV operations

Given the electricity consumption and emission factor of the power sector across different regions, the carbon emissions released by BEV operations across three climate zones can be calculated by

$$X_{MLRYR} = \eta_{MLRYR} \cdot E_{MLRYR} \quad (8)$$
$$X_{North} = \eta_{North} \cdot E_{North} \quad (9)$$
$$X_{South} = \eta_{South} \cdot E_{South} \quad (10)$$

The total carbon emissions released by the best-selling BEV operations in China (X) can be expressed as follows:

$$X = X_{MLRYR} + X_{North} + X_{South} \quad (11)$$

## 3. RESULTS

### 3.1 Trends in energy and emissions of best-selling BEV model operations in the MLRYR of China

Fig. 1 outlines the electricity consumption and associated carbon emissions of the top-20 selling BEV models in China's MLRYR from 2020 to 2022. From 2020 to 2022, electricity consumption for the top-20 BEVs in China's MLRYR rose from 155.7 GWh to 858.2 GWh. Tesla Model 3 and Wuling Hongguang Mini EV dominated in 2020, while Tesla Model Y led from 2021 onwards. BYD models also held significant shares. Carbon emissions from these models rose from 0.08 $MtCO_2$ in 2020 to 0.42 $MtCO_2$ in 2022, mirroring the growth in electricity use. Despite the increasing demand for BEVs, the environmental impact depends on the electricity grid's emission factor, highlighting the need for cleaner energy sources.

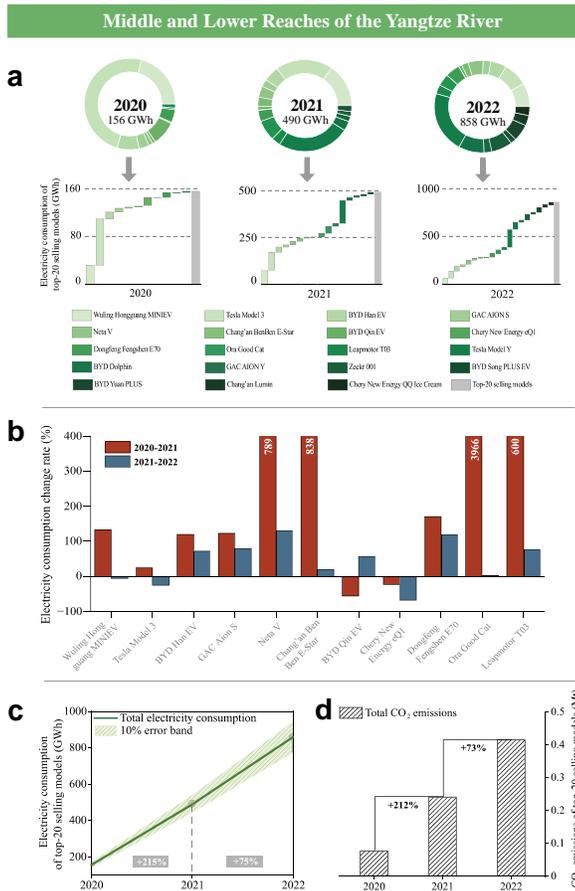

Fig. 1 Electricity consumption of the top-20 selling BEV model operations in the MLRYR of China, 2020-2022: (a) contribution of each model to total electricity consumption, (b) annual change rate of electricity use in the key models, and (c) trends in total electricity consumption. (d) Carbon emissions released by the top-20 selling BEV model operations in the MLRYR of China, 2020-2022.

### 3.2 Trends in energy and emissions of best-selling BEV model operations in North China

As shown in Fig. 2, over the period from 2020 to 2022, electricity consumption demonstrated a remarkable growth trajectory. Specifically, in 2020, electricity consumption totaled 187.0 GWh. Subsequently, it surged by 149.0% in 2021, reaching 465.52 GWh, and further increased by 63.4% in 2022, totaling 760.9 GWh. These data highlight a distinct pattern of expansion, which is particularly noteworthy during the period from 2021 to 2022, where despite a slightly moderated growth rate, a relatively high rate of increase was maintained. Concurrently, the associated carbon emissions witnessed a substantial upsurge in tandem with the rise in electricity consumption. In 2020, carbon emissions reached 0.13 $MtCO_2$. After experiencing a remarkable growth rate of 144.5% by 2021, the total emissions reached 0.33 $MtCO_2$. Subsequently, in 2022, despite a deceleration in growth rate, they continued to rise at a rate of 61.0%, culminating in a total of 0.53 $MtCO_2$.

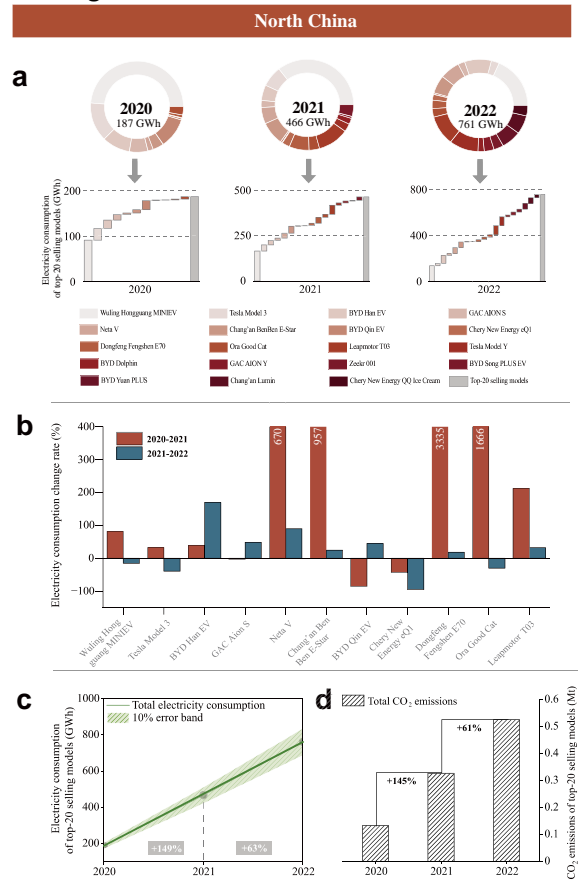

Fig. 2 Electricity consumption of the top-20 selling BEV model operations in North China, 2020-2022: (a)

contribution of each model to total electricity consumption, (b) annual change rate of electricity use in the key models, and (c) trends in total electricity consumption. (d) Carbon emissions released by the top-20 selling BEV model operations in North China, 2020-2022.

### 3.3 Trends in energy and emissions of best-selling BEV model operations in South China

As shown in Fig. 3, the total electricity consumption of the top-20 selling models in South China in 2020 was 258.57 GWh, 1.4 times that of North China and 1.7 times that of the MLRYR during the same year. Moreover, South China experienced the fastest growth in electricity use from 2020 to 2021, reaching 229.0%. From 2021 to 2022, electricity consumption in South China increased from 850.7 GWh to 1,434.5 GWh. Although the growth in electricity consumption in South China slowed from 2021 to 2022, it still maintained a relatively high growth rate of 68.6%. Notably, during this period, the growth rate of electricity consumption in the MLRYR exceeded that in South China, becoming the fastest-growing region in terms of electricity consumption.

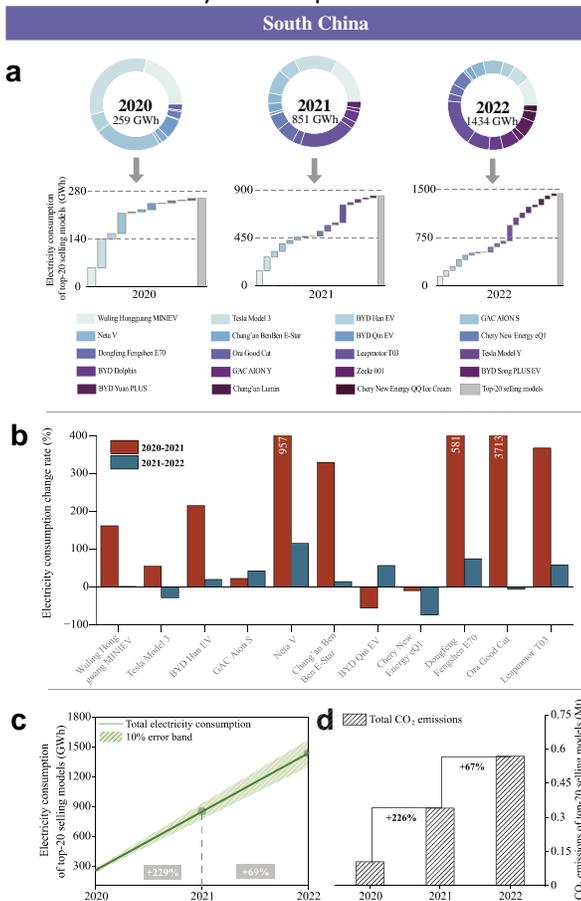

Fig. 3 Electricity consumption of the top-20 selling BEV model operations in South China, 2020-2022: (a) contribution of each model to total electricity consumption, (b) annual change rate of electricity use in the key models, and (c) trends in total electricity consumption. (d) Carbon emissions released by the top-20 selling BEV model operations in South China, 2020-2022.

## 4. DISCUSSION

### 4.1 Energy and emission features of nationwide best-selling BEV model operations

Fig. 4 presents the electricity consumption and carbon emissions for China's top-20 best-selling BEV models from 2020 to 2022. In 2020, electricity consumption was 601.2 GWh (±10%), while in 2021, it surged to 1806.5 GWh, a 200.5% increase. By 2022, consumption reached 3053.6 GWh, reflecting a 69.0% growth rate from the previous year. This rise is due to factors like increased BEV numbers and changing driving patterns. Carbon emissions also rose from 2020 to 2022, going from 0.35 $MtCO_2$ to 1.04 $MtCO_2$ and then to 1.73 $MtCO_2$. The increase from 2020 to 2021 was 195.4%, and from 2021 to 2022, it slowed to 66.9%, indicating a stabilization in growth. Although electricity consumption and carbon emissions are strongly correlated, the smaller growth in emissions compared to electricity use suggests progress in decarbonizing China's power sector.

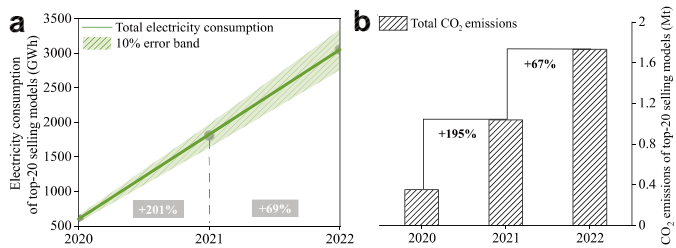

Fig. 4 (a) Total electricity consumption and (b) associated $CO_2$ emissions of the top-20 selling BEV operations in China, 2020-2022.

### 4.2 Energy and emission intensities of the best-selling BEV model operations

From 2020 to 2022, electricity consumption intensity in China's North, South, and the Middle-Lower Yangtze River (MLRYR) regions showed a general decline, with North China consistently having the highest intensity. In 2020, North China's electricity consumption intensity was 1796.1 kWh/vehicle, significantly higher than MLRYR's 1209.1 kWh/vehicle and South China's 1243.5 kWh/vehicle. By 2021, the gap narrowed, with North China's intensity dropping to 1446.1 kWh/vehicle. South China saw the fastest decline from 2021 to 2022, at - 11.3%, reaching 1024.9 kWh/vehicle. Carbon emission intensity followed a similar pattern, with North China showing the highest intensity, nearly double that of the MLRYR and South China. However, from 2020 to 2022, all regions showed a decline in carbon emissions, with North China experiencing the sharpest drop between 2020 and 2021 (-20.9%). Notable models like the Wuling Hongguang MINI EV and Tesla Model 3 had relatively lower electricity consumption intensities across the regions. While some models showed increases in intensity, the general trend was toward reduced electricity consumption and carbon emissions, indicating ongoing efforts to decarbonize China's power sector.

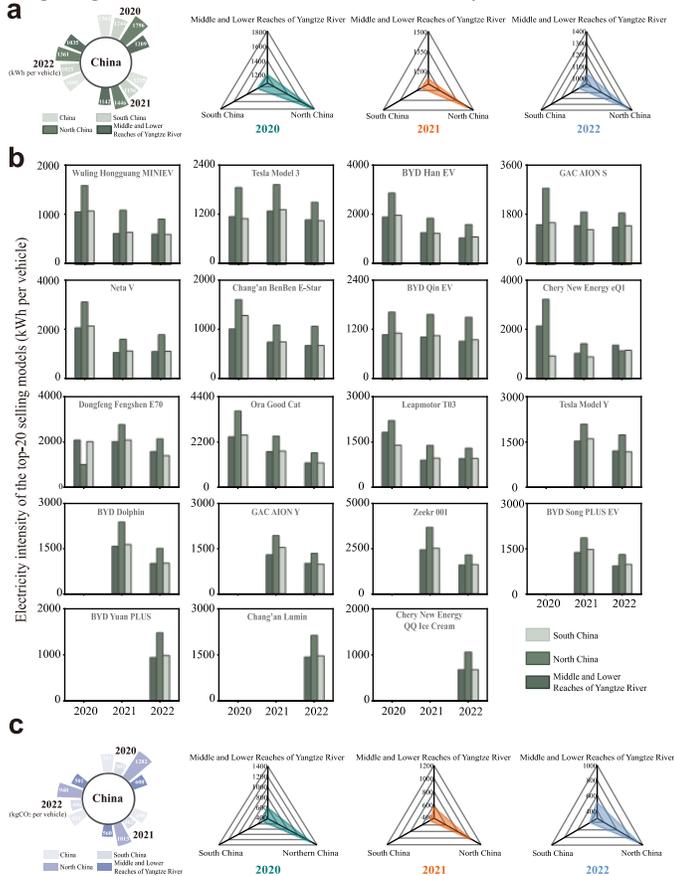

Fig. 5 (a) Average electricity use intensity and (c) average carbon intensity of the best-selling BEVs' operations; (b) electricity use intensity of the top-20 selling models across different climate zones in China, 2020-2022.

### 4.3 Comparison of energy and emissions between the best-selling BEV models and BEV stocks

This study used a standardized bottom-up charging model to estimate energy and emissions for the top-selling BEV models, providing a way to scale these results to the entire national BEV stock. This scaling method can estimate total electricity consumption and carbon emissions for all BEVs in China. Ratios of national BEV stock to top-20 BEV sales were 7.9, 4.2, and 4.0 from 2020 to 2022, with corresponding ratios for all BEV sales to top-20 sales at 2.1, 1.8, and 1.6. Fig. 6 shows that estimates from BEV stock are nearly double those from BEV sales data. In 2022, the total electricity consumption of China's BEV stock was 12,048 GWh, with 6.8 MtCO$_2$ in carbon emissions. The gap between top-20 sales and full-scale estimates narrowed over time, partly because some best-selling BEVs were introduced in 2021 and 2022.

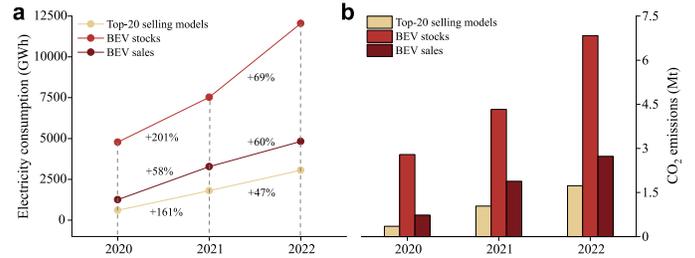

Fig. 6 Trends in (a) electricity consumption and (b) associated carbon emissions for the top-20 selling BEV models, BEV sales, and BEV stocks in China, 2020-2022.

## 5. CONCLUSIONS

This study established a bottom-up charging demand model to measure the electricity usage and associated carbon emissions of best-selling BEV models across various climate zones in China since 2020. Specifically, the charging demand of each BEV model, including its various versions, was assessed by considering parameters such as vehicle model performance, sales, mileage, climate conditions, driver behaviors, and the emission factor of the power sector. Additionally, the energy and emission intensities of best-selling BEV model operations, along with the energy demand and associated emissions of BEV stocks in China's passenger car sector, were analyzed. The principal findings are summarized as follows:

- The operational energy demand of the top-20 selling BEV models in China increased from 601 GWh to 3054 GWh during 2020-2022, with BEVs in South China contributing approximately 50% of the total electricity demand.
- From 2020 to 2022, the energy and carbon intensities of the top-selling models decreased from 1364 to 1095 kWh per vehicle and 797 to 621 kgCO$_2$ per vehicle, respectively, with North China experiencing the highest intensity decline compared to that in other regions.
- The operational energy demand of BEV stocks in China increased from 4774 GWh to 12,048 GWh during 2020-2022, while the associated carbon

emissions of BEV stocks rose to 6.8 MtCO$_2$ in 2022, reflecting an annual growth rate exceeding 50%.

This study has uncovered a few questions that require further exploration, pointing toward promising directions for future studies. It is crucial to scrutinize the interplay between BEV carbon emissions and the development of residential charging systems, encompassing the nuances in microgrids and building power systems across diverse zones. Incorporating these facets into upcoming studies will enhance the breadth and depth of insights into the analysis of energy use and associated carbon emissions for EVs at both the national and regional levels.

**ACKNOWLEDGEMENT**

This manuscript was authored by an author at Lawrence Berkeley National Laboratory under Contract No. DE-AC02-05CH11231 with the U.S. Department of Energy. The U.S. Government retains, and the publisher, by accepting the article for publication, acknowledges, that the U.S. Government retains a non-exclusive, paid-up, irrevocable, world-wide license to publish or reproduce the published form of this manuscript, or allows others to do so, for U.S. Government purposes.

**REFERENCE**
[1] Yan R, Ma M, Zhou N, Feng W, Xiang X, Mao C. Towards COP27: Decarbonization patterns of residential building in China and India. Applied Energy. 2023;352:122003.
[2] Pan Y, Fang W, Zhang W. Development of an energy consumption prediction model for battery electric vehicles in real-world driving: A combined approach of short-trip segment division and deep learning. Journal of Cleaner Production. 2023;400:136742.
[3] Yuan H, Ma M, Zhou N, Xie H, Ma Z, Xiang X, Ma X. Battery electric vehicle charging in China: Energy demand and emissions trends in the 2020s. Applied Energy 2024;365:123153.
[4] Ahmed I, Adnan M, Ali M, Kaddoum G. Supertwisting sliding mode controller for grid-to-vehicle and vehicle-to-grid battery electric vehicle charger. Journal of Energy Storage. 2023;70:107914.
[5] Xiang X, Zhou N, Ma M, Feng W, Yan R. Global transition of operational carbon in residential buildings since the millennium. Advances in Applied Energy 2023;11:100145.
[6] He H, Sun F, Wang Z, Lin C, Zhang C, Xiong R, et al. China's battery electric vehicles lead the world: achievements in technology system architecture and technological breakthroughs. Green Energy and Intelligent Transportation. 2022;1:100020.
[7] Zahoor A, Yu Y, Zhang H, Nihed B, Afrane S, Peng S, et al. Can the new energy vehicles (NEVs) and power battery industry help China to meet the carbon neutrality goal before 2060? Journal of Environmental Management. 2023;336:117663.
[8] Yan R, Chen M, Xiang X, Feng W, Ma M. Heterogeneity or illusion? Track the carbon Kuznets curve of global residential building operations. Applied Energy 2023;347:121441.
[9] Lee G, Song J, Han J, Lim Y, Park S. Study on energy consumption characteristics of passenger electric vehicle according to the regenerative braking stages during real-world driving conditions. Energy. 2023;283:128745.
[10] Li K, Ma M, Xiang X, Feng W, Ma Z, Cai W, Ma X. Carbon reduction in commercial building operations: A provincial retrospection in China. Applied Energy 2022;306:118098.
[11] Zhang S, Zhou N, Feng W, Ma M, Xiang X, You K. Pathway for decarbonizing residential building operations in the US and China beyond the mid-century. Applied Energy. 2023;342:121164.
[12] Ma M, Pan T, Ma Z. Examining the Driving Factors of Chinese Commercial Building Energy Consumption from 2000 to 2015: A STIRPAT Model Approach. Journal of Engineering Science & Technology Review 2017;10:28-34.
[13] Ma M, Yan R, Cai W. A STIRPAT model-based methodology for calculating energy savings in China's existing civil buildings from 2001 to 2015. Natural Hazards 2017;87:1765-1781.
[14] Yan R, Ma M, Pan T. Estimating energy savings in Chinese residential buildings from 2001 to 2015: A decomposition analysis. Journal of Engineering Science & Technology Review 2017;10:107-113.
[15] Ma M, Yan R, Cai W. Energy savings evaluation in public building sector during the 10th–12th FYP periods of China: an extended LMDI model approach. Natural Hazards 2018;92:429-441.
[16] Zhang S, Xiang X, Ma Z, Ma M, Zou C. Carbon Neutral Roadmap of Commercial Building Operations by Mid-Century: Lessons from China. Buildings 2021;11:510.